\documentclass[final,5p,times,twocolumn]{elsarticle}
\usepackage{amsmath,amssymb,bm}
\usepackage{xcolor}
\usepackage[colorlinks=true,citecolor=blue,linkcolor=blue,urlcolor=blue]{hyperref}
\biboptions{numbers,sort&compress}
\newcommand{\dd}{\mathrm{d}}
\journal{Physics Letters B}

\usepackage{stfloats}
\usepackage{balance}
\usepackage{placeins}
\usepackage{afterpage}
\usepackage{tikz}
\definecolor{lime}{HTML}{A6CE39}

\DeclareRobustCommand\orcidJonathan{{\href{https://orcid.org/0000-0001-9291-0893}{\orcidicon}}}
\DeclareRobustCommand\orcidEdson{{\href{https://orcid.org/0000-0001-9929-5977}{\orcidicon}}}
\DeclareRobustCommand\orcidRM{{\href{https://orcid.org/0000-0003-3123-059X}{\orcidicon}}}

\makeatletter
\newcommand{\authorEmail}[2]{\begingroup\@eadauthor={#1}\ead{#2}\endgroup}
\makeatother
\begin{document}
\begin{frontmatter}
\title{Hyperbolic scaling of the small-exponent limit in magnetic nonlinear electrodynamics}
\author[ifce]{Jonathan A. Rebou\c{c}as\,\orcidJonathan\hspace{2mm}}
\authorEmail{Jonathan A. Rebou\c{c}as}{jalvesreboucas@ifce.edu.br}
\author[UC]{R. Martínez von Dossow\,\orcidRM\hspace{2mm}}
\authorEmail{R. Martínez von Dossow}{rmartinezvd@uc.cl}
\author[ufca]{Edson Otoniel\,\orcidEdson\hspace{2mm}}
\authorEmail{Edson Otoniel}{edson.otoniel@ufca.edu.br}
\address[UC]{Facultad de F\'isica, Pontificia Universidad Cat\'olica de Chile, Vicu\~na Mackenna 4860, Santiago, Chile}
\address[ifce]{Instituto Federal de Educa\c{c}\~ao, Ci\^encia e Tecnologia do Cear\'a (IFCE), Iguatu, CE, Brazil}
\address[ufca]{Universidade Federal do Cariri (UFCA), Instituto de Forma\c{c}\~ao de Educadores - IFE, R. Oleg\'ario Emidio de Araujo S/N, Brejo Santo - CE, 63.260-000 - Brazil}
\begin{abstract}
A magnetic spacetime can approach Schwarzschild while its electromagnetic characteristics remain singular in the same parameter limit. We study this separation in fractional-power nonlinear electrodynamics as the exponent tends to zero with asymptotic mass, magnetic charge and nonlinear coupling independently fixed. The extraordinary polarization admits a common physical time direction only outside a boundary that grows as the inverse fourth root of the exponent. On this radial scale, the lapse tends to unity but the constitutive angular response retains a nondegenerate barrier and finite scaled critical data. An exact representation of the exterior electromagnetic energy yields complete second-order corrections to the orbit and impact, separating mass and charge contributions and their distinct cancellations. Constraining the mass to equal the electromagnetic self-energy preserves the leading critical constants while changing the approach from fourth-root to square-root corrections. Returning rays remain inside the admissible exterior and define a limiting scattering function with a calculated logarithmic strong-deflection coefficient. In the weak regime, constitutive bending decays as the inverse fourth power of scaled impact, whereas finite mass supplies a slower inverse-impact tail. We determine their competition scale, its accessibility above the critical impact, and the accuracy of the two-term approximation. Independent energy, orbit and scattering calculations check the expansions and resolve their parameter-dependent convergence. The analysis shows how a nearly vacuum background can support a distinct exterior optical limit through the nonlinear constitutive response.
\end{abstract}
\end{frontmatter}
\section{Introduction}\label{sec:introduction}
Nonlinear electrodynamics makes the relation between a gravitating electromagnetic field and its optical response more intricate than in Maxwell theory. The historical field-theory programme of Born and Infeld linked particle properties and electromagnetic energy, while subsequent gravitational constructions established magnetic black holes and monopoles as settings in which nonlinear matter determines the geometry \cite{BornInfeld1934,Bronnikov2001}. A limit of the background metric consequently raises a separate question: does the propagation of electromagnetic disturbances approach the same optical behaviour?

Regular electrically charged solutions and the nonlinear magnetic interpretation of the Bardeen model illustrate how changing the electromagnetic source can alter the central geometry \cite{AyonBeatoGarcia1998,AyonBeatoGarcia2000}. Electric constructions with a de Sitter center also require attention to the effective geometry experienced by electromagnetic disturbances \cite{Dymnikova2004}. These examples motivate a distinction between metric regularity and optical admissibility. The present family addresses an exterior singular limit with an independently specified asymptotic mass; regularity at the center is not imposed.

The relevant propagation law follows from the characteristics of the field equations. For nonlinear electromagnetic Lagrangians, constitutive derivatives enter the wavefront equation and can produce polarization-dependent optical geometries \cite{Boillat1970,Novello,ObukhovRubilar2002}. The distinction between the background and characteristic cones is explicit even in Born--Infeld theory, where fluctuations are governed by the Boillat geometry \cite{GibbonsHerdeiro2001}. Small corrections to the background lapse therefore need not imply small corrections to the angular response experienced by a selected polarization.

A characteristic geometry must also possess an admissible propagation domain. Causality and unitarity constrain the electromagnetic response through derivatives of the Lagrangian, and symmetric hyperbolicity is tied to a compatible intersection of the effective cones \cite{ShabadUsov2011,Abalos2015}. Analyses of strong-field causality further show that nonlinear variants require restrictions beyond their formal resemblance to Born--Infeld theory \cite{BornAgain2024}. These considerations make the common physical time direction part of the optical problem, rather than a condition that can be inferred from the gravitational horizon alone.

Admissibility of characteristics and dynamical stability answer different questions. A coupled perturbation analysis allows fluctuations of both the metric and the electromagnetic field, with stability conditions depending on the constitutive law \cite{MorenoSarbach2003}. The instability rate of a circular ray supplies a local optical timescale, but cannot replace that field-theoretic analysis. We use it only to interpret the near-critical scattering coefficient.

The fractional-power theory introduced by Kruglov provides a concrete family in which this distinction becomes singular \cite{KruglovNotes,Kruglov}. Its magnetic solutions have an exact mass function and a continuous nonlinear exponent. We use the sector with no coupling to the second electromagnetic invariant and specify the asymptotic mass independently of the conserved magnetic charge and nonlinear coupling. This choice separates the central mass integration constant from the electromagnetic self-mass. Requiring the entire mass to be electromagnetic instead defines a different trajectory through the same constitutive family.

Effective null geodesics and weak scattering in Kruglov and related nonlinear models have already been investigated \cite{Habibina}. Ray-traced Born--Infeld images likewise illustrate the optical consequences of a magnetic nonlinear response \cite{He2022}. More directly, Ramadhan and collaborators studied photon orbits, deflection and imaging in Kruglov electrodynamics, including the outward displacement of the outer orbit as the exponent decreases \cite{Ramadhan}. The present question concerns the controlled scale and corrections of that small-exponent behaviour. We retain returning extraordinary rays in the connected admissible exterior; their critical impact is an orbital and scattering quantity, with image formation requiring its own radiation prescription \cite{Gralla2019}.

The geometrical description of trapped light extends beyond the Schwarzschild photon sphere through the notion of photon surfaces \cite{Claudel2001}. Light rings and fundamental photon orbits organize a wider range of strong-lensing phenomena \cite{CunhaHerdeiro2018}; nonlinear electromagnetic black holes with several photon spheres can support multiple critical curves and higher-order images \cite{Hui2025}. Here the object to be continued is the outer nondegenerate maximum in a specified connected characteristic exterior. Its uniqueness in the limiting problem does not imply uniqueness for arbitrary nonlinear electromagnetic backgrounds. Analytical shadow constructions distinguish the critical impact from the angular size seen by an observer \cite{PerlickTsupko2022}, reinforcing the need to keep the present scattering problem separate from a complete image model.

Large Schwarzschild bending already produces sequences of relativistic images \cite{VirbhadraEllis2000}. Exact spherical lens equations connect ray integrals and image observables without a weak-deflection approximation \cite{Perlick2004}. In the weak regime, optical-metric methods relate bending to Gaussian curvature through the Gauss--Bonnet theorem \cite{GibbonsWerner2008}. We instead evaluate the exterior ray integral directly, using strong- and weak-deflection expansions as separately controlled descriptions of that same integral.

Strong-deflection methods supply the local logarithmic description near a nondegenerate photon orbit, including ultrastatic limiting geometries \cite{Bozza2002,Tsukamoto2017}. Circular-orbit instability and its relation to lensing coefficients provide a complementary interpretation of the corresponding timescales \cite{Cardoso2009,Stefanov2010}. Those frameworks establish the general mechanism. They do not, by themselves, determine the radial scaling, mass-boundary dependence or competition of weak-bending contributions in a singular constitutive limit.

We address these issues by separating convergence at fixed physical radius from the characteristic problem on a growing radial scale. An exact representation of the exterior electromagnetic energy gives the complete second-order corrections to the critical orbit and impact, while an independently constructed electromagnetic-mass path identifies how the approach rate changes with the mass condition. Exterior scattering then resolves both the limiting strong-deflection coefficient and a weak-bending transition between constitutive and mass contributions. The transition is assessed together with its accessibility to returning rays and the finite-exponent accuracy of the approximation.

This paper is organized as follows. Section~\ref{sec:theory} specifies the magnetic theory and mass family. Section~\ref{sec:characteristics} derives the polarization cones and their admissible exterior. Section~\ref{sec:limits} constructs the scaled limit and electromagnetic tail. Section~\ref{sec:orbit} develops the critical-orbit corrections and compares mass conditions. Section~\ref{sec:scattering} treats returning rays and the strong- and weak-deflection regimes. Section~\ref{sec:conclusions} summarizes the resulting physical picture.

\section{Magnetic theory and the independent-mass family}\label{sec:theory}
We consider a single-invariant nonlinear electromagnetic field coupled to general relativity, with action
\begin{equation}
 S=\frac{1}{16\pi}\int\dd^4x\sqrt{-g}\,[R+4\mathcal L_\nu(\mathcal F)].
 \label{eq:action}
\end{equation}
Here $R$ is the Ricci scalar, $g$ the metric determinant, and $\mathcal F=F_{\mu\nu}F^{\mu\nu}/4$ the electromagnetic invariant. We use signature $(-,+,+,+)$ and $G=c=1$. The fractional-power Lagrangian is
\begin{equation}
 \mathcal L_\nu(\mathcal F)=\frac{1-(1+\beta\mathcal F/\nu)^\nu}{\beta}.
 \label{eq:lagrangian}
\end{equation}
The exponent $\nu$ is dimensionless and $\beta>0$ has dimensions of length squared in these geometrized field units; $\mathcal F$ and $\mathcal L_\nu$ have dimensions of inverse length squared. This is the zero-second-invariant-coupling sector of Kruglov's theory \cite{Kruglov}, rather than the full two-invariant Born--Infeld theory. Removing that coupling is a constitutive assumption and matters for polarization. The small-exponent analysis uses $0<\nu<1/2$.

For a magnetic monopole, $F_{\theta\phi}=Q\sin\theta$ and $\mathcal F=Q^2/(2r^4)$, where $Q>0$ has dimensions of length. The static spherical line element is
\begin{equation}
 \dd s^2=-f_\nu(r)\dd t^2+\frac{\dd r^2}{f_\nu(r)}+r^2\dd\Omega^2.
 \label{eq:metric}
\end{equation}
The coordinate $r$ is areal radius, $t$ is normalized at infinity, and $\dd\Omega^2$ is the metric of the unit sphere. Write $f_\nu=1-2m_\nu/r$. The Einstein equation and asymptotic mass condition give
\begin{equation}
 m_\nu'(r)=-r^2\mathcal L_\nu,\qquad
 m_\nu(r)=M-\int_r^\infty s^2[-\mathcal L_\nu(s)]\dd s.
 \label{eq:mass_equation}
\end{equation}
The constant $M>0$ is the mass at infinity, with dimensions of length. We hold $M,Q,\beta$ independently fixed while varying $\nu$. A central mass integration constant is therefore allowed. Imposing a vanishing central integration constant would select a different path and change the gravitational corrections derived below.

The energy integral yields
\begin{equation}
 \begin{split}
 f_\nu(r)={}&1-\frac{2M}{r}-\frac{2r^2}{3\beta}\left[\vphantom{\frac11}
 {}_2F_1\!\left(-\frac34,-\nu;\frac14;
 -\frac{\beta Q^2}{2\nu r^4}\right)-1\right].
 \end{split}\label{eq:lapse}
\end{equation}
Here ${}_2F_1$ is the Gauss hypergeometric function. At fixed nonzero $\nu$, the large-radius expansion is $f_\nu=1-2M/r+Q^2/r^2+O(r^{-6})$. In particular, the charge term has the Reissner--Nordstr\"om sign and normalization. At $\nu=1$, Eq.~\eqref{eq:lagrangian} becomes exactly Maxwell's theory and Eq.~\eqref{eq:lapse} becomes the exact Reissner--Nordstr\"om lapse. This is an algebraic control outside the small-exponent range, not an extrapolation of its asymptotic formulas.

Introduce the constitutive length
\begin{equation}
 \ell_\nu=\left(\frac{\beta Q^2}{2\nu}\right)^{1/4}.
 \label{eq:scale}
\end{equation}
It grows as $\nu^{-1/4}$ although the conserved charge stays fixed. The total electromagnetic self-mass is finite for $\nu<3/4$ and equals
\begin{equation}
 m_{\rm em}=\frac{\nu\ell_\nu^3}{3\beta}
 B\!\left(\frac14,\frac34-\nu\right),
 \label{eq:self_mass}
\end{equation}
where $B$ is Euler's beta function. To obtain this expression, rescale the energy integral by $r=\ell_\nu y$, integrate the factor $y^2[(1+y^{-4})^\nu-1]$ by parts, and substitute $y^4$ in the remaining integral. Both boundary terms vanish when $\nu<3/4$. Thus $m_{\rm em}=O(\nu^{1/4})$ on the adopted parameter path.

Since the energy density is nonnegative, $0\leq M-m_\nu(r)\leq m_{\rm em}$. Consequently,
\begin{equation}
 0\leq f_\nu(r)-\left(1-\frac{2M}{r}\right)
 \leq\frac{2m_{\rm em}}{r}.
 \label{eq:fixed_bound}
\end{equation}
The convergence to Schwarzschild is uniform on each fixed interval bounded away from $r=0$. For sufficiently small $\nu$, $M>m_{\rm em}$ and the central mass is positive. The lapse then diverges negatively at the center and becomes positive at infinity, so at least one gravitational horizon exists. This statement does not require the characteristic field to remain admissible down to that horizon.

\section{Polarizations and the characteristic exterior}\label{sec:characteristics}
\subsection{The principal symbol and its normalization}
Electromagnetic characteristics can differ from the background metric cone \cite{Novello,He2022}. Their normalization can be established directly from the field equation $\nabla_\mu(\mathcal L_{\mathcal F}F^{\mu\nu})=0$. Let $k_\mu$ be a discontinuity covector and write the leading field discontinuity as $k_\mu a_\nu-k_\nu a_\mu$, where $a_\mu$ is a polarization amplitude. Define $v^\nu=k_\mu F^{\mu\nu}$ and $k^2=g^{\mu\nu}k_\mu k_\nu$. In the gauge $k\cdot a=0$, the principal equation is
\begin{equation}
 \mathcal L_{\mathcal F}k^2a^\nu+
 \mathcal L_{\mathcal F\mathcal F}v^\nu(v\cdot a)=0.
 \label{eq:principal}
\end{equation}
One physical polarization has $v\cdot a=0$ and propagates on $k^2=0$. The extraordinary polarization instead satisfies $\mathcal L_{\mathcal F}k^2+\mathcal L_{\mathcal F\mathcal F}v^2=0$. The gauge mode is not a third polarization. For radial propagation the two physical cones coincide; the distinction concerns their response to tangential momentum.

For Eq.~\eqref{eq:lagrangian}, with the dimensionless field ratio $X=\beta\mathcal F/\nu$, differentiation gives
\begin{equation}
 \mathcal L_{\mathcal F}=-(1+X)^{\nu-1},\qquad
 \mathcal L_{\mathcal F\mathcal F}=
 \frac{\beta(1-\nu)}{\nu}(1+X)^{\nu-2}.
 \label{eq:derivatives}
\end{equation}
The first derivative is strictly negative and nonzero for the finite magnetic fields considered here. A conformal representative of the extraordinary characteristic metric is
\begin{equation}
 \dd s_{\rm e}^2=-f_\nu\dd t^2+\frac{\dd r^2}{f_\nu}
 +\frac{r^2}{H_\nu(r)}\dd\Omega^2,
 \label{eq:effective_metric}
\end{equation}
where the dimensionless angular response is
\begin{equation}
 H_\nu=1+\frac{2\mathcal F\mathcal L_{\mathcal F\mathcal F}}
 {\mathcal L_{\mathcal F}}
 =\frac{1+(2\nu-1)X}{1+X}.
 \label{eq:cone}
\end{equation}
The factor of two follows from the quarter-contracted invariant. With $I=F_{\mu\nu}F^{\mu\nu}=4\mathcal F$, the equivalent Lagrangian obeys $\mathcal L_I=\mathcal L_{\mathcal F}/4$ and $\mathcal L_{II}=\mathcal L_{\mathcal F\mathcal F}/16$. Hence the customary factor-four tensor written using derivatives with respect to $I$ gives precisely Eq.~\eqref{eq:principal} after conversion.

Ramadhan et al. already studied Kruglov photon orbits and deflection, including the outward escape of the outer orbit as the exponent decreases \cite{Ramadhan}. In version 3, Eqs.~(2), (9) and (11), their printed effective tensor uses a factor four alongside a quarter-contracted invariant and yields an angular denominator proportional to $1+(8\nu-7)X$. That expression is not the characteristic tensor obtained from Eq.~\eqref{eq:action}. We therefore compare the established qualitative mechanism, while deriving the cone and all numerical constants in the convention above. Their orbit and image curves are not numerical validation data for this branch. Related nonlinear photon-orbit studies likewise establish the broader context \cite{Habibina}; they do not fix the coefficients of the present singular limit.

\subsection{A common time direction}
In the static background orthonormal frame, the extraordinary dispersion relation is
\begin{equation}
 \omega^2=k_{\hat r}^2+H_\nu k_\perp^2.
 \label{eq:dispersion}
\end{equation}
Here $\omega$ is frequency, $k_{\hat r}$ is radial wave number, and $k_\perp$ is the tangential wave-number magnitude. Static-time propagation of arbitrary real spatial wave vectors requires $H_\nu>0$. In the adopted exponent range, $0<H_\nu\leq1$ on this branch. Its strict inner boundary is
\begin{equation}
 r_{\rm h}=\ell_\nu(1-2\nu)^{1/4}.
 \label{eq:boundary}
\end{equation}
The subscript h denotes hyperbolicity of this electromagnetic branch, whereas $r_H$ below denotes a gravitational horizon. The magnetic causality and unitarity bound in Appendix A, Eq.~(34), of Kruglov~\cite{Kruglov} gives the same radial restriction after substituting the monopole field. The boundary itself is therefore an established constraint; here its growing scale is related to the singular-limit orbit and exterior scattering. At $r_{\rm h}$ the principal angular part degenerates. Below it, transverse wave vectors can have imaginary static-frame frequency. A sign change is not a proof that the full four-dimensional quadratic form becomes Euclidean; the common physical time direction is the relevant criterion.

The same condition follows from a local constitutive calculation. In a magnetic frame with field magnitude $\mathcal B$ and direction $\hat z$, $\mathcal F=\mathcal B^2/2$. The electric permittivity is $-\mathcal L_{\mathcal F}$, while the inverse magnetic permeability has transverse entries $-\mathcal L_{\mathcal F}$ and longitudinal entry $-\mathcal L_{\mathcal F}-2\mathcal F\mathcal L_{\mathcal F\mathcal F}$. The plane-wave Maxwell system then has squared physical phase speeds $1$ and $\cos^2\vartheta+H_\nu\sin^2\vartheta$, where $\vartheta$ is the angle between wave vector and background field. Numerical eigenvalues of that independent local system agree with Eq.~\eqref{eq:dispersion} to below $5\times10^{-16}$ in the sampled cases.

We retain the connected static region extending to infinity with $f_\nu>0$ and $r>r_{\rm h}$. At small exponent, the boundary moves far outside the gravitational horizon. It supplies neither an absorbing surface nor a continuation rule for the electromagnetic field. Exterior circular orbits and rays that turn before reaching it are nevertheless well-defined characteristic problems.

\section{The singular radial limit and the electromagnetic tail}\label{sec:limits}
\subsection{An exact scaled representation}
Define the dimensionless variables and fixed charge-to-mass ratio
\begin{equation}
 y=\frac{r}{\ell_\nu},\qquad \delta=\frac{M}{\ell_\nu},\qquad
 \chi=\frac{Q^2}{M^2}.
 \label{eq:scaled_variables}
\end{equation}
Then $\nu=\beta Q^2\delta^4/(2M^4)$ and $Q^2/\ell_\nu^2=\chi\delta^2$. The angular response is exactly $H_\nu(y)=(y^4-1+2\nu)/(y^4+1)$. Its limit at fixed $y>1$ is
\begin{equation}
 h(y)=\frac{y^4-1}{y^4+1}.
 \label{eq:limiting_cone}
\end{equation}
Although $\mathcal F=\nu/(\beta y^4)$ vanishes in this limit, the ratio $\beta\mathcal F/\nu$ stays finite. Constitutive derivatives thus retain a finite angular response. Taking a weak-field limit at fixed exponent before changing $\nu$ would discard it. At fixed physical radius, instead, $X$ diverges and $H_\nu\to-1$; every such point eventually lies outside the adopted extraordinary domain.

The energy integral in Eq.~\eqref{eq:mass_equation} gives an exact scaled lapse,
\begin{equation}
 f_\nu(\ell_\nu y)=1-\frac{2\delta}{y}
 +\chi\delta^2 A_\nu(y),
 \label{eq:exact_scaled_lapse}
\end{equation}
where the dimensionless positive tail function is
\begin{equation}
 A_\nu(y)=\frac1y\int_y^\infty t^2
 \frac{\exp[\nu\ln(1+t^{-4})]-1}{\nu}\dd t.
 \label{eq:tail_nu}
\end{equation}
The exterior field energy makes the lapse larger than the mass-only expression at the same $M$. This term is essential at order $\delta^2$. It is not represented by adjusting the integration constant or by fitting a correction to the critical radius.

For $y$ bounded away from zero, Taylor expansion under the integral gives $A_\nu(y)=A(y)+O(\nu)$, with
\begin{equation}
 A(y)=\frac1y\int_y^\infty t^2\ln(1+t^{-4})\dd t.
 \label{eq:tail_limit}
\end{equation}
Indeed, the remainder in the exponential is bounded by a constant times $\nu\ln^2(1+t^{-4})$, and its product with $t^2$ is integrable at infinity. On any compact $y$ interval in $y>1$, this estimate also holds after the finite number of derivatives needed below. Integration by parts and differentiation give
\begin{equation}
 \begin{split}
 A(y)&=-\frac{y^2}{3}\ln(1+y^{-4})
 +\frac{4}{3y}\int_y^\infty\frac{t^2}{1+t^4}\dd t,\\
 A'(y)&=-y\ln(1+y^{-4})-\frac{A(y)}y.
 \end{split}\label{eq:tail_identities}
\end{equation}
The tail decreases with radius, and $A(y)=y^{-2}+O(y^{-6})$ at large $y$. The leading term recovers the scaled Reissner--Nordstr\"om contribution. Equations~\eqref{eq:tail_nu} and \eqref{eq:lapse} were evaluated independently at 55-digit precision; the largest lapse difference at the stated test points was below $1.3\times10^{-51}$. This checks both the sign and the normalization of the term that competes with the second-order orbit displacement.

\begin{figure*}[!tbp]
\begin{minipage}{\textwidth}
\centering\includegraphics[width=\textwidth]{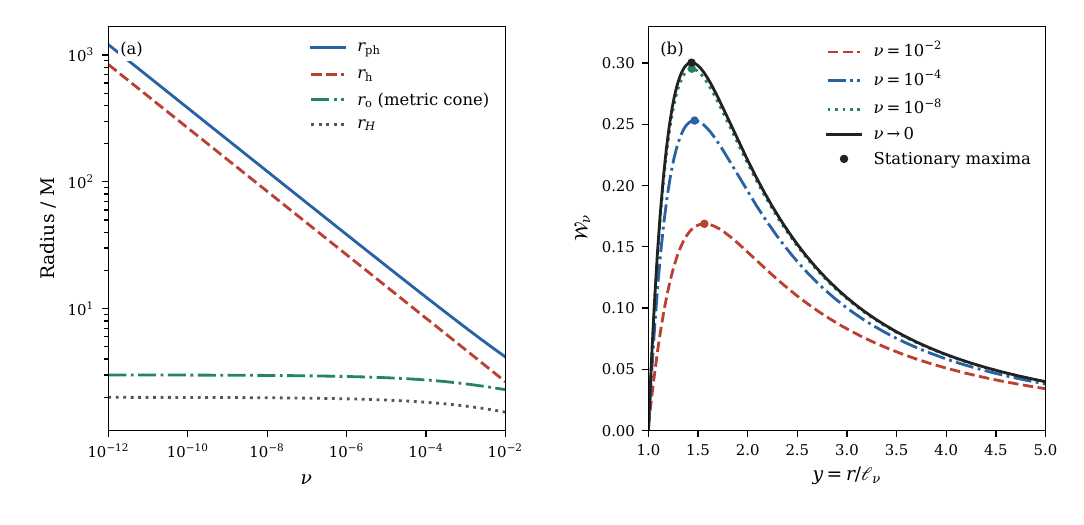}
\caption{Separation of scales and the extraordinary barrier at $M=Q=\beta=1$. (a) Exact critical radius $r_{\rm ph}$, characteristic boundary $r_{\rm h}$, metric-cone orbit $r_{\rm o}$ and gravitational horizon $r_H$, over $10^{-12}\leq\nu\leq10^{-2}$. (b) Exact scaled potentials for $\nu=10^{-2},10^{-4},10^{-8}$ and their limit, on $1\leq y\leq5$; dots mark their maxima, identified by the stationary-maxima legend entry with the same family colors. This interval lies in the strict finite-exponent exterior. The metric-cone orbit is a formal comparison where it lies below $r_{\rm h}$; no interior optical continuation is assumed.}
\label{fig:geometry}
\end{minipage}
\end{figure*}

\subsection{A finite barrier in a weakly curved region}
Let $E$ and $L$ denote the conserved extraordinary-ray energy and angular momentum, and let $b=L/E$ be the impact measured at infinity. A dot denotes an affine derivative for the conformal representative in Eq.~\eqref{eq:effective_metric}. The radial and angular equations are
\begin{equation}
 \dot r^2=E^2[1-b^2U_\nu(r)],\qquad
 \dot\phi=\frac{L H_\nu}{r^2},\qquad
 U_\nu=\frac{f_\nu H_\nu}{r^2}.
 \label{eq:ray_equations}
\end{equation}
The coordinate $\phi$ is the equatorial azimuth. The normalization of the affine parameter is irrelevant to the path. Since $f_\nu,H_\nu\to1$ at infinity for every finite $\nu$, $b$ is the usual asymptotic impact, with dimensions of length.

Define the dimensionless potential $\mathcal W_\nu(y)=\ell_\nu^2U_\nu(\ell_\nu y)$. Its leading shape is
\begin{equation}
 w(y)=\frac{h(y)}{y^2}=\frac{y^4-1}{y^2(y^4+1)}.
 \label{eq:limiting_potential}
\end{equation}
Combining Eqs.~\eqref{eq:cone} and \eqref{eq:exact_scaled_lapse}, the fixed-coupling expansion becomes
\begin{equation}
 \mathcal W_\nu(y)=w(y)+\delta v_1(y)
 +\chi\delta^2v_2(y)+O(\delta^4),
 \label{eq:potential_expansion}
\end{equation}
where $v_1=-2w/y$ and $v_2=wA$ are dimensionless. There is no explicit order-$\delta^3$ term in this potential. The first correction to $H_\nu$ is order $\nu=O(\delta^4)$, whereas the tail correction to $A$ first contributes at order $\delta^6$. The estimate is local on compact intervals in $y>1$, at fixed finite $M,Q,\beta$; it does not extend uniformly to arbitrary charge ratios or to a moving critical-impact endpoint.

Figure~\ref{fig:geometry} connects the growing characteristic scale to its finite barrier. The extraordinary orbit and its admissibility boundary move outward together, whereas the gravitational horizon and metric-cone orbit approach $2M$ and $3M$. Once the latter orbit lies below $r_{\rm h}$, it does not describe propagation in a region where the full electromagnetic system shares a common physical time direction. In the admissible exterior, the barrier approaches $w$ while its maximum shifts with the slower order-$\nu^{1/4}$ gravitational correction. The surviving barrier therefore originates in the constitutive angular response even though the lapse tends to one on this scale.

\section{Critical orbit through second order}\label{sec:orbit}
\subsection{The limiting maximum and its continuation}
The potential $w$ is positive for $y>1$ and vanishes at the two ends of that interval. Its stationary equation reduces to $y^8-4y^4-1=0$, giving a unique exterior maximum at
\begin{equation}
 y_*=(2+\sqrt5)^{1/4},\qquad
 B_*=[w(y_*)]^{-1/2}.
 \label{eq:critical_constants}
\end{equation}
Here $B_*$ is the limiting impact in units of $\ell_\nu$. Numerically, $y_*=1.4346327151$ and $B_*=1.8248810035$. Its curvature is $w''(y_*)=-16\sqrt5/(3+\sqrt5)^2<0$, so the maximum is nondegenerate. The implicit-function theorem continues it to a unique nearby maximum at sufficiently small $\delta$. This establishes the branch relevant to the limit, without asserting that every finite-exponent geometry has only one stationary point globally.

Write the continued critical data as $y_c=r_{\rm ph}/\ell_\nu$ and $B_c=b_{\rm ph}/\ell_\nu$. With primes denoting $y$ derivatives and a star indicating evaluation at $y_*$, perturbing $\mathcal W_\nu'(y_c)=0$ yields
\begin{equation}
 y_c=y_*+a\delta+(c_M+\chi c_Q)\delta^2+O(\delta^3),
 \label{eq:radius_second}
\end{equation}
where the first coefficient is $a=-v_{1*}'/w_*''=1/(2\sqrt5)$. The second-order coefficients separate the two physical contributions:
\begin{equation}
 c_M=-\frac{\tfrac12w_*'''a^2+v_{1*}''a}{w_*''},
 \qquad c_Q=-\frac{v_{2*}'}{w_*''}.
 \label{eq:radius_coefficients}
\end{equation}
The first arises from moving the maximum in the mass-perturbed potential. The second comes from the exterior electromagnetic energy. Since $v_{2*}'=w_*A_*'<0$ and $w_*''<0$, $c_Q<0$: the field tail partially compensates the positive mass-induced displacement. Keeping $M$ independent of the field self-mass is necessary for this separation.

The critical height gives
\begin{equation}
 \frac{B_c}{B_*}=1+\frac{\delta}{y_*}
 +(e_M+\chi e_Q)\delta^2+O(\delta^3),
 \label{eq:impact_second}
\end{equation}
with
\begin{equation}
 e_M=\frac{3}{2y_*^2}+\frac{a^2w_*''}{4w_*},
 \qquad e_Q=-\frac{A_*}{2}.
 \label{eq:impact_coefficients}
\end{equation}
The order-$\delta$ orbit displacement cancels from the height at first order because $w_*'=0$. At second order its curvature contribution remains, together with the electromagnetic tail. Thus the radius and the critical impact probe related but different combinations of the same perturbations.

\begin{table}[!tbp]
\caption{Dimensionless coefficients calculated from the limiting potential and tail integral. They are analytic evaluations, not fitted quantities. The separate mass and charge terms combine with $\chi=Q^2/M^2$ in Eqs.~\eqref{eq:radius_second} and \eqref{eq:impact_second}.}
\label{tab:coefficients}
\centering\small
\begin{tabular}{lrr}
\hline
Quantity & Mass term & Electromagnetic term\\
\hline
Radius & $c_M=0.387818895$ & $c_Q=-0.146200562$\\
Impact/$B_*$ & $e_M=0.674480683$ & $e_Q=-0.237646862$\\
\hline
\end{tabular}
\end{table}

In physical units, Eq.~\eqref{eq:radius_second} contains the divergent term $\ell_\nu y_*$, the finite correction $aM$, and an order-$M^2/\ell_\nu$ correction with both mass and charge contributions. The corresponding impact expansion has finite correction $B_*M/y_*$. These finite offsets can remain significant in a numerical sequence long after the leading exponent $-1/4$ is apparent.

\subsection{Charge-dependent compensation and convergence}
The coefficients in Table~\ref{tab:coefficients} vanish at different charge ratios. The radius coefficient changes sign at $\chi=-c_M/c_Q\simeq2.652650$, whereas the relative-impact coefficient vanishes at $\chi=-e_M/e_Q\simeq2.838164$. These are cancellations of second-order terms on a specified parameter path, not critical charges of the background or exact identities at finite exponent. The first-order mass terms remain positive in both quantities. At either cancellation, the corresponding residual starts at order $\delta^3$ unless a further independent cancellation occurs.

To expose the complete second-order terms, subtract the leading and first-order values and divide by $\delta^2$:
\begin{equation}
 R_r=\frac{y_c-y_*-a\delta}{\delta^2},\qquad
 R_b=\frac{B_c/B_*-1-\delta/y_*}{\delta^2}.
 \label{eq:compensated}
\end{equation}
Their limits are $c_M+\chi c_Q$ and $e_M+\chi e_Q$. In particular, a mass-only second-order approximation would approach the wrong nonzero plateau for every fixed $Q>0$.

\begin{figure}[!tbp]
\begin{minipage}{\columnwidth}
\centering\includegraphics[width=\columnwidth]{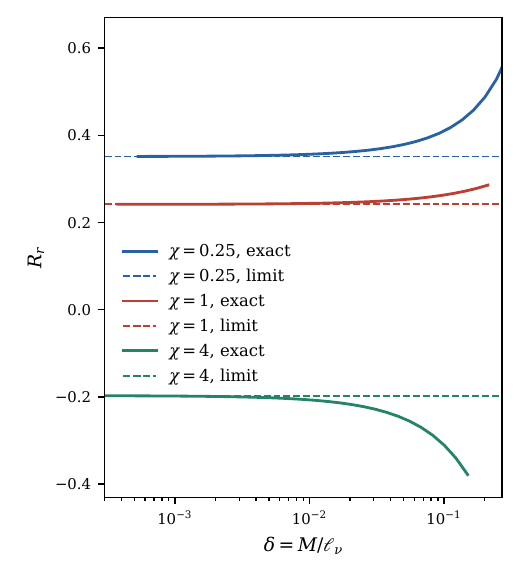}
\caption{Compensated second-order radius correction $R_r$ at $M=\beta=1$, for $Q=0.5,1,2$ and $10^{-14}\leq\nu\leq10^{-3}$. Solid curves are exact; dashed horizontal levels are $c_M+\chi c_Q$. Each color and line type is identified in the legend. Each family uses its own $\delta$; $3\times10^{-4}\leq\delta\leq0.27$ is displayed. Impact coefficients are given analytically in Eq.~\eqref{eq:impact_coefficients} and Table~\ref{tab:coefficients}.}
\label{fig:charge}
\end{minipage}
\end{figure}
Figure~\ref{fig:charge} separates charge effects that the leading divergent scale suppresses. The three families approach distinct second-order plateaus, including negative corrections at $\chi=4$. This sign belongs to the subtracted term; it does not reverse the positive first-order displacement or the divergence of the physical radius and impact. The complete electromagnetic tail is needed to obtain these plateaus. For the impact, expanding the height of the shifted maximum and its inverse square root already gives the coefficients in Eq.~\eqref{eq:impact_coefficients}. Those coefficients retain an independent physical distinction: their cancellation charge differs from the radius cancellation charge. The impact result is therefore retained analytically and in Table~\ref{tab:coefficients}, without repeating the plateau plot.

To represent the accuracy of both critical quantities in one diagnostic, define
\begin{equation}
 \epsilon_{\max}=\max\left\{
 \frac{|y_c-y_{\rm app}|}{y_c},
 \frac{|B_c-B_{\rm app}|}{B_c}\right\}.
 \label{eq:joint_error}
\end{equation}
Both individual relative errors are bounded by this maximum. Thus a curve below a chosen tolerance guarantees that both approximations meet it at that sampled point; no equality of their finite-exponent errors is assumed.

\begin{figure}[!tbp]
\begin{minipage}{\columnwidth}
\centering\includegraphics[width=\columnwidth]{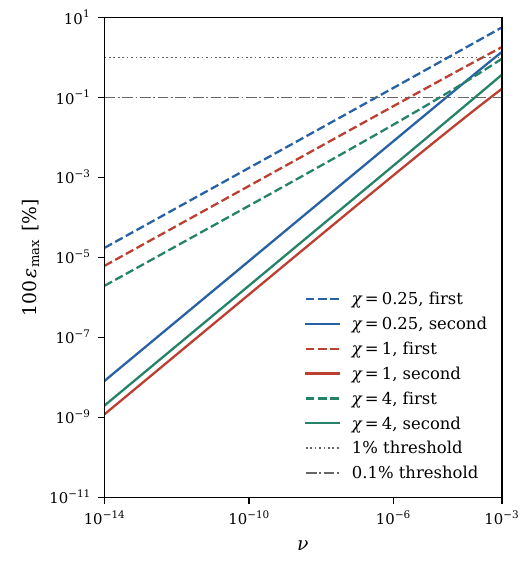}
\caption{Joint accuracy of radius and impact for the same fixed-mass families as Fig.~\ref{fig:charge}. The plotted quantity is $100\epsilon_{\max}$ from Eq.~\eqref{eq:joint_error}, in percent. For each charge, dashed and solid curves retain first and complete second order. The gray dotted and dash-dotted levels indicate $1\%$ and $0.1\%$. All curves and thresholds are identified in the legend. No coefficients are fitted.}
\label{fig:residuals}
\end{minipage}
\end{figure}
Figure~\ref{fig:residuals} shows the gain from second order through the larger of the two relative errors, preserving the more restrictive accuracy requirement at each exponent. The asymptotic absolute remainders are $O(\nu^{1/2})$ at first order and $O(\nu^{3/4})$ at second order; their relative errors have the same limiting powers because the scaled critical data tend to nonzero constants. At a finite exponent, however, a small remainder is not guaranteed by its power alone. Table~\ref{tab:accuracy} quantifies this difference at the upper endpoint.

\begin{table}[!tbp]
\caption{Relative second-order errors, in percent, at $\nu=10^{-3}$ with $M=\beta=1$. Each error is normalized by its own exact critical quantity; their maximum is plotted in Fig.~\ref{fig:residuals}.}
\label{tab:accuracy}\centering\small
\begin{tabular}{crr}\hline
$\chi$ & Radius error [\%] & Impact error [\%]\\\hline
$0.25$ & $1.3517$ & $1.3965$\\
$1$ & $0.1318$ & $0.1668$\\
$4$ & $0.2773$ & $0.3761$\\\hline
\end{tabular}
\end{table}
Critical roots satisfy $f_\nu,H_\nu>0$ and have negative curvature. A separate inverse-radius series calculation reproduces the critical data and their compensated corrections. High-precision roots give a second-order residual slope close to $3/4$ in $\nu$ at $M=Q=\beta=1$; details of arithmetic and convergence are retained with the numerical data. The asymptotic window also depends on the fixed couplings: for $\beta/M^2=100$ and five sampled charge ratios, a cubic slope in $\delta$ is approached more closely at $\delta=10^{-5}$--$10^{-4}$ than at $10^{-3}$--$10^{-2}$. This is consistent with the fixed-coupling, rather than uniform-in-coupling, remainder in Eq.~\eqref{eq:potential_expansion}.

\begin{figure}[!tbp]
\begin{minipage}{\columnwidth}
\centering\includegraphics[width=\columnwidth]{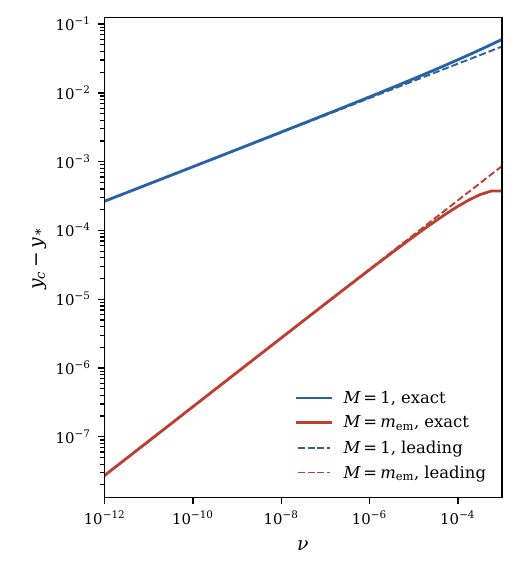}
\caption{Critical-radius displacement at $Q=\beta=1$, comparing $M=1$ with $M=m_{\rm em}(\nu)$ over $10^{-12}\leq\nu\leq10^{-3}$. Solid curves are exact and dashed curves are the calculated leading corrections from Eqs.~\eqref{eq:radius_second} and \eqref{eq:selfmass_path_critical}. The legend identifies both paths and both approximation types. The impact rates follow analytically from Eq.~\eqref{eq:path_ratio}.}
\label{fig:mass_path}
\end{minipage}
\end{figure}

\subsection{Contrast with a self-mass-constrained path}\label{sec:mass_path}
The asymptotic mass condition is a physical part of the limit. As a controlled contrast, keep $Q,\beta$ fixed but set $M=m_{\rm em}(\nu)$, thereby removing the central mass integration constant. The mass now decreases with the exponent, and the previously fixed ratio $\chi$ diverges. Substitution into a fixed-$\chi$ remainder would therefore be unjustified. The exact scaled lapse instead provides a direct expansion on this second path.

Introduce $\mu=Q^2/\ell_\nu^2=O(\nu^{1/2})$ and $\mathcal B_\nu=B(1/4,3/4-\nu)$. Equation~\eqref{eq:self_mass} implies $\delta=\mu\mathcal B_\nu/6$, so
\begin{equation}
 f_\nu(\ell_\nu y)=1+\mu\left[A_\nu(y)-\frac{\mathcal B_\nu}{3y}\right].
 \label{eq:selfmass_path_lapse}
\end{equation}
The gravitational and exterior-energy terms now enter at the same leading order. Let $J(y)=A(y)-\mathcal B_0/(3y)$, with $\mathcal B_0=B(1/4,3/4)$. On compact $y>1$ intervals, $\mathcal W_\nu=w+\mu wJ+O(\mu^2)$; the cone correction is order $\nu=O(\mu^2)$. Perturbing the nondegenerate maximum gives
\begin{equation}
 \begin{split}
 y_c&=y_*-\mu\frac{w_*J_*'}{w_*''}+O(\mu^2),\\
 \frac{B_c}{B_*}&=1-\frac{\mu J_*}{2}+O(\mu^2).
 \end{split}\label{eq:selfmass_path_critical}
\end{equation}
For $Q=\beta=1$, the two order-$\mu$ coefficients are approximately $0.01937591$ and $0.27849952$. The approach to the same limiting orbit is therefore of order $\nu^{1/2}$, rather than the order-$\nu^{1/4}$ correction of the fixed-$M$ family. The leading constitutive constants survive because $f_\nu\to1$ on the growing scale in both cases. Their corrections retain the choice of mass boundary condition.

The similar leading powers of the radius and impact do not require separate convergence plots. Let $\Delta y=y_c-y_*$ and $\Delta B=B_c/B_*-1$. Dividing the nonzero leading terms in Eqs.~\eqref{eq:radius_second}, \eqref{eq:impact_second} and \eqref{eq:selfmass_path_critical} gives
\begin{equation}
 \lim_{\nu\to0^+}\frac{\Delta B}{\Delta y}=
 \begin{cases}
 (a y_*)^{-1}, & M\ \text{fixed},\\[2pt]
 \dfrac{J_*w_*''}{2w_*J_*'}, & M=m_{\rm em}(\nu).
 \end{cases}
 \label{eq:path_ratio}
\end{equation}
Here $Q,\beta$ remain fixed in both paths. The first ratio has an $O(\delta)$ correction; the second has an $O(\mu)$ correction. Their limits are finite and positive, so the impact inherits the radius exponent, $1/4$ or $1/2$ respectively. This statement concerns leading rates, not an identity between the full finite-exponent curves.

Figure~\ref{fig:mass_path} displays the radius corrections for the two mass conditions without dividing out their powers; Eq.~\eqref{eq:path_ratio} supplies the corresponding impact-rate check. At small exponent, the fixed-mass correction decays as $\nu^{1/4}$ and the self-mass correction as $\nu^{1/2}$. Both lead to the same scaled critical constants, but their finite-exponent departures differ in size and slope. The disagreement with the dashed leading self-mass radius correction near the upper endpoint also shows why compensation alone is not a measure of finite-exponent accuracy. This comparison concerns exterior orbits; an electromagnetic mass relation alone does not establish a horizon for every member.

\begin{figure*}[!tbp]
\begin{minipage}{\textwidth}
\centering\includegraphics[width=\textwidth]{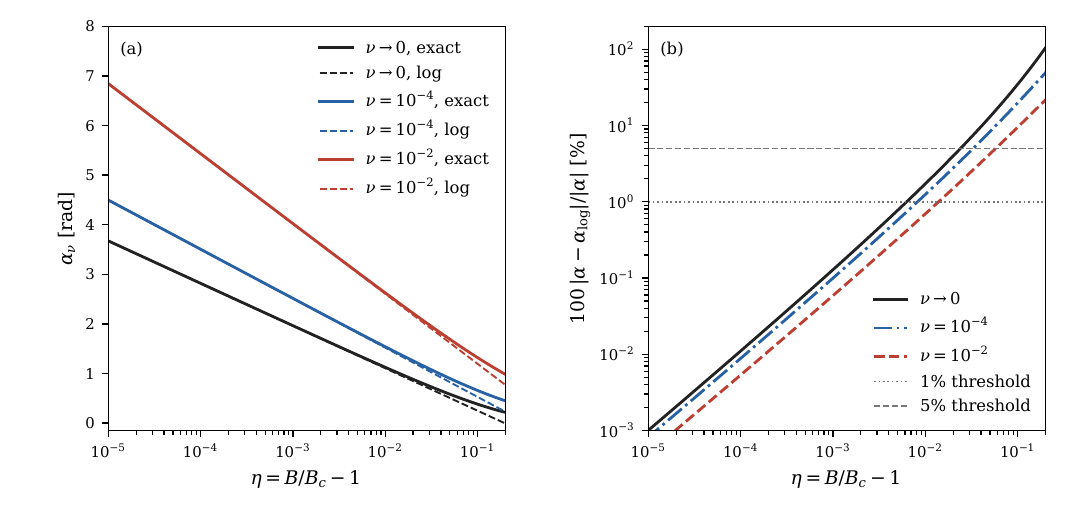}
\caption{Exact exterior deflection and its logarithmic approximation for $M=Q=\beta=1$, at $\nu=10^{-2},10^{-4}$ and in the scaled limit. (a) Exact curves are solid and logarithmic approximations dashed; the legend identifies each family and approximation separately. Each family uses its own $\eta=B/B_c-1$. (b) Relative errors, normalized by the exact angle, with gray dotted $1\%$ and dashed $5\%$ thresholds, both identified in the legend. The displayed range is $10^{-5}\leq\eta\leq0.2$. The constants are calculated independently of these error curves; a negative extrapolated approximation near the far endpoint signals loss of validity, not a negative exact deflection.}
\label{fig:strong}
\end{minipage}
\end{figure*}

\section{Returning rays in the admissible exterior}\label{sec:scattering}
\subsection{An exterior scattering function}
For an incoming extraordinary ray with $b>b_{\rm ph}$ on the small-exponent branch, let $r_t>r_{\rm ph}$ denote its outer turning radius. Its path returns to the same asymptotic region and never reaches the characteristic boundary. The full azimuth minus the straight-line angle is
\begin{equation}
 \alpha_\nu(B)=2\int_{y_t}^\infty
 \frac{B H_\nu(y)\,\dd y}{y^2\sqrt{1-B^2\mathcal W_\nu(y)}}-\pi,
 \label{eq:deflection}
\end{equation}
where $B=b/\ell_\nu$ and $y_t=r_t/\ell_\nu$ satisfy $B^2\mathcal W_\nu(y_t)=1$. Every symbol in this integral refers to the characteristic branch; replacing its numerator by the background metric-cone expression would change the result even when the lapse is almost flat.

At fixed $B>B_*$, sufficiently separated from criticality, the limiting deflection is obtained by replacing $H_\nu,\mathcal W_\nu$ with $h,w$. This defines $\alpha_*(B)$. The outer turning root is simple and its limit lies strictly beyond $y_*$. A squared-distance change of variable removes its square-root endpoint singularity. The large-$y$ integrand is bounded by a constant times $B/y^2$, allowing the limiting integral to be taken after separately controlling its tail. The argument fails when the impact approaches criticality at the same time as $\nu$; that case needs its own estimate.

The numerical integral uses $u=1/y$ followed by $u=u_t(1-s^2)$, which regularizes the turning endpoint and includes infinity. A local derivative expansion avoids cancellation in the potential difference. Independent calculations use a convergent inverse-radius energy series with $u=u_t\sin\theta$, and an areal-radius integration for limiting rays. They reproduce the exact curves and the flat-space control; the Schwarzschild comparison below checks a separate strong-deflection problem.

\subsection{Strong deflection and orbital instability}
Near a nondegenerate optical maximum, the logarithmic mechanism is established in spherical lensing \cite{Bozza2002,Tsukamoto2017}. For fixed exponent and $\eta=B/B_c-1\to0^+$, Eq.~\eqref{eq:deflection} gives
\begin{equation}
 \alpha_\nu(B)=-\bar a_\nu\ln\eta+\bar b_\nu+o(1),
 \label{eq:strong_deflection}
\end{equation}
where $\bar b_\nu$ is the finite part of the exterior integral. Expanding the potential quadratically around $y_c$ yields the coefficient
\begin{equation}
 \bar a_\nu=\frac{H_\nu(y_c)}{y_c^2}
 \sqrt{\frac{2}{-\mathcal W_\nu''(y_c)}}.
 \label{eq:abar}
\end{equation}
The numerator is the angular weight on a single radial segment. The factor of two for incoming and outgoing propagation is already included in Eq.~\eqref{eq:strong_deflection}: the turning displacement scales as $\sqrt{\eta}$, converting the two radial logarithms into one $\ln\eta$.

In the scaled exponent limit, this becomes
\begin{equation}
 \bar a_*^2=\frac{\sqrt5-1}{4\sqrt5}.
 \label{eq:abar_limit}
\end{equation}
Thus $\bar a_*\simeq0.3717480345$, rather than the Schwarzschild coefficient one. This is a model-specific evaluation of the usual nondegenerate-maximum mechanism. The logarithm itself and strong lensing in Kruglov backgrounds are established results, not contributions claimed here \cite{Ramadhan}. The finite part $\bar b_\nu$ is evaluated numerically; no closed expression for it is assumed.

The same coefficient can be related to the orbit's local instability. Let $\Omega_c=\dd\phi/\dd t$ at the orbit and let $\lambda_c>0$ be the exponential growth rate of a small radial displacement in the asymptotically normalized time $t$. The radial equation gives
\begin{equation}
 \ell_\nu\Omega_c=\frac1{B_c},\qquad
 \ell_\nu\lambda_c=f_\nu(y_c)
 \sqrt{-\frac{\mathcal W_\nu''(y_c)}{2\mathcal W_\nu(y_c)}}.
 \label{eq:frequencies}
\end{equation}
Equations~\eqref{eq:abar} and \eqref{eq:frequencies} imply $\bar a_\nu=\Omega_c/\lambda_c$. Both frequencies vanish in physical units as the orbit escapes, but their ratio tends to the finite constant in Eq.~\eqref{eq:abar_limit}. A flat background limit therefore does not force the extraordinary winding coefficient to its ordinary Schwarzschild value. This radial-orbit instability does not establish stability of the electromagnetic or gravitational background.

Figure~\ref{fig:strong} makes the local character of strong deflection quantitative. The logarithmic approximation follows the exact curve near criticality, while its error increases as the turning point moves away from the maximum. Aligning the endpoints by each family's $B_c$ retains differences in the barrier shape and winding coefficient. The plotted error is an approximation error, not a quadrature uncertainty. Table~\ref{tab:strong} gives the limiting and finite-exponent coefficients; convergence checks on their numerical finite parts are independent of the displayed comparison.

\begin{table}[!tbp]
\caption{Critical impacts and strong-deflection coefficients at $M=Q=\beta=1$. The first row denotes the scaled limit, not the original Lagrangian at zero exponent. The finite parts are numerical estimates.}
\label{tab:strong}\centering\small
\begin{tabular}{crrr}\hline
$\nu$ & $B_c$ & $\bar a_\nu$ & $\bar b_\nu$\\\hline
$\to0$ & $1.824881$ & $0.371748$ & $-0.607668$\\
$10^{-4}$ & $1.988082$ & $0.430932$ & $-0.466068$\\
$10^{-2}$ & $2.434557$ & $0.613394$ & $-0.219547$\\\hline
\end{tabular}
\end{table}
The finite parts in Table~\ref{tab:strong} are evaluated at turning offsets $y_t-y_c=10^{-6},3\times10^{-7},10^{-7}$ with 55-digit arithmetic. Reducing the offset across this sequence changes the finite part by less than $9\times10^{-12}$ radians in each family. This is a convergence diagnostic at finite offsets, rather than a rigorous truncation bound. The six printed decimal places are stable under this check.

For an independent gravitational benchmark, replace the constitutive response by $H=1$ and take the Schwarzschild lapse in the same quadrature. The known finite part is $\ln[216(7-4\sqrt3)]-\pi$ and the logarithmic coefficient is one \cite{Bozza2002}. At the turning radius $r_t=3.001M$, the numerical finite part differs by less than $9.3\times10^{-7}$ radians. This is a check against a different, analytically controlled scattering problem. It does not reconcile the invariant-convention difference with the Kruglov comparison identified in Sec.~\ref{sec:characteristics}.

\begin{figure*}[!tbp]
\begin{minipage}{\textwidth}
\centering\includegraphics[width=\textwidth]{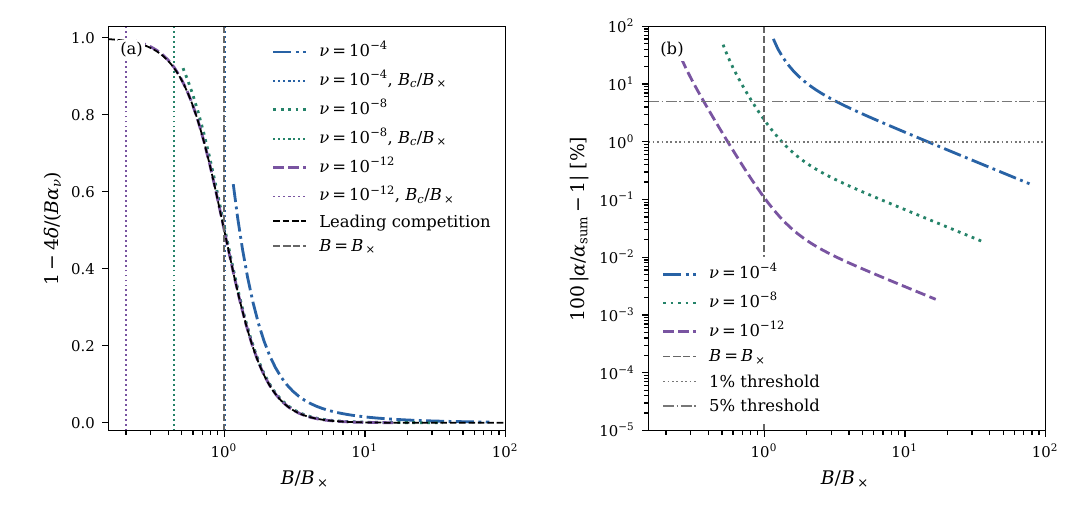}
\caption{Weak-bending competition and accuracy at $M=Q=\beta=1$, using exact returning rays with $2\leq y_t\leq150$. (a) Mass-subtracted ratio $1-4\delta/(B\alpha_\nu)$ and the leading prediction $[1+(B/B_\times)^3]^{-1}$ (black dashed). Colored vertical dotted lines locate each family's $B_c/B_\times$; its returning rays require larger impact. (b) Relative error of the two-term sum, normalized by that sum, with gray dotted $1\%$ and dash-dotted $5\%$ thresholds. The legends identify every critical boundary, equality line and threshold. The vertical gray dashed line is $B=B_\times$. All sampled errors are included. The $\nu=10^{-4}$ family is a control outside the accessible weak-crossover regime, as quantified in Table~\ref{tab:crossover}. The subtracted ratio is not a flux fraction.}
\label{fig:crossover}
\end{minipage}
\end{figure*}

\subsection{Weak bending and the competing large-impact scales}
The limiting geometry has unit lapse and spatial circumferential radius $\mathcal R=y/\sqrt{h(y)}$. At large $y$, $\mathcal R=y+y^{-3}+O(y^{-7})$, hence $\dd y/\dd\mathcal R=1+3\mathcal R^{-4}+O(\mathcal R^{-8})$. Expressing the limiting angular integral in this radius and substituting $\mathcal R=B/\sin\psi$ gives
\begin{equation}
 \alpha_*(B)=\frac{9\pi}{8B^4}+O(B^{-8}).
 \label{eq:weak_limit}
\end{equation}
The coefficient follows from $2\times3\int_0^{\pi/2}\sin^4\psi\,\dd\psi=9\pi/8$. This route uses the circumferential-radius deformation, rather than a fit to small numerical angles. At an outer turning radius $y_t=100$, the independent quadrature recovers the leading coefficient with relative error below $6\times10^{-8}$, as expected from the next inverse-impact order.

For small but finite $\delta$, the metric mass produces the slower weak-bending term $4\delta/B$. Expanding the exact angular integral at large $B$, with fixed finite $\chi$ and the fixed-coupling relation $\nu=O(\delta^4)$, gives the two leading contributions
\begin{equation}
 \begin{split}
 \alpha_\nu(B)={}&\frac{4\delta}{B}+\frac{9\pi}{8B^4}\\
 &+O\!\left(\frac{\delta^2}{B^2}+\frac{\delta}{B^5}
 +\frac1{B^8}+\frac{\nu}{B^4}\right).
 \end{split}\label{eq:weak_composite}
\end{equation}
The first error includes the mass-squared and charge corrections. To see the ordering, at large $y$ the lapse starts with $1-2\delta/y+\chi\delta^2/y^2$, while the angular factor starts with $1-2(1-\nu)/y^4$. Removing the simple turning root before expanding makes the weak-field integrals regular. Their first mass and constitutive terms give Eq.~\eqref{eq:weak_composite}; mixed and quadratic terms have the displayed orders. This expansion concerns large impacts, separately from the critical expansion in Eq.~\eqref{eq:strong_deflection}.

The two terms are equal at the dimensionless scale
\begin{equation}
 B_\times=\left(\frac{9\pi}{32\delta}\right)^{1/3}.
 \label{eq:crossover}
\end{equation}
An equality scale corresponds to a returning ray only if
\begin{equation}
 B_\times>B_c.
 \label{eq:crossover_domain}
\end{equation}
Weak bending additionally requires large impact and separation from the critical endpoint; Eq.~\eqref{eq:crossover_domain} alone is insufficient. Table~\ref{tab:crossover} evaluates both accessibility and the approximation at equality. The quantities $B_c$ and $B_\times$ are calculated separately.

For sufficiently small $\delta$, the intermediate window $1\ll B\ll B_\times$ is dominated by the constitutive contribution. At $B\gg B_\times$ the mass term dominates. Taking $\delta\to0$ first leaves the $B^{-4}$ law; taking $B\to\infty$ first at any nonzero $\delta$ selects the $B^{-1}$ law. This is another nonuniform limit, now entirely within the admissible returning-ray family.

Figure~\ref{fig:crossover} distinguishes equality of two asymptotic terms from an accessible weak-scattering transition. At $\nu=10^{-4}$, $B_\times<B_c$, so no returning ray can reach the estimated equality scale. This control also has a maximum displayed sum error of about $60.3\%$ near its smallest turning radius. The two smaller exponents have an accessible equality scale, and their errors there decrease as that scale separates from the critical barrier. The two-term expression becomes increasingly accurate toward large impact, but the constitutive term then becomes subdominant; accurate total bending alone does not imply equally accurate isolation of that smaller contribution.

\begin{table}[!tbp]
\caption{Accessibility and two-term error at $B=B_\times$ for $M=Q=\beta=1$. Error means $100(\alpha_\nu/\alpha_{\rm sum}-1)$, where $\alpha_{\rm sum}$ is the displayed sum in Eq.~\eqref{eq:weak_composite}. The absent entry has no returning ray at equality.}
\label{tab:crossover}\centering\small
\begin{tabular}{crr}\hline
$\nu$ & $B_\times/B_c$ & Error at equality [\%]\\\hline
$10^{-4}$ & $0.98151$ & ---\\
$10^{-8}$ & $2.28463$ & $2.3886$\\
$10^{-12}$ & $4.95907$ & $0.10743$\\\hline
\end{tabular}
\end{table}
At $B=X B_\times$ with fixed $X>0$ and $\delta\to0$, the leading sum is $O(\delta^{4/3})$ and the leading error is $O(\delta^{8/3})$. Thus its relative error is $O(\delta^{4/3})=O(\nu^{1/3})$ on compact $X$ intervals away from zero, once those intervals lie above the critical barrier. Independent quadratures at $X=1$ confirm this order. These estimates concern weak scattering; they do not provide a simultaneous exponent and near-critical error theorem.

All scattering calculations above terminate at an outer turning point. Subcritical rays, a shadow boundary, emitted intensity and polarization-dependent flux require additional propagation and radiation assumptions inside the characteristic exterior. No absorption law is assigned to $r_{\rm h}$, and the present exterior critical impact is not identified with a measured shadow \cite{Gralla2019}.

\section{Conclusions}\label{sec:conclusions}
The small-exponent optical limit is governed by the region in which the electromagnetic field retains an admissible characteristic evolution. For the extraordinary polarization studied here, that region moves outward even as the magnetic background approaches Schwarzschild at fixed radius. Rescaling with the constitutive length exposes a finite barrier in a region whose lapse tends to unity. Its critical orbit is therefore controlled by the angular electromagnetic response rather than by a finite curvature correction near the Schwarzschild photon sphere.

The exterior field energy determines the complete second-order displacement of this orbit. Its charge-dependent contribution partially compensates the mass contribution, with different cancellations for the radius and critical impact. The leading scaled constants survive a change from independently fixed mass to electromagnetic self-mass, but the approach changes from the fourth root to the square root of the exponent. The analytic ratios between the radius and impact corrections establish the shared rate within each path, while the joint error diagnostic measures the accuracy of both approximations at finite exponent.

Returning rays provide a scattering problem wholly inside the characteristic exterior. Their near-critical winding has a finite limiting logarithmic coefficient, whose interpretation follows from the ratio of orbital frequency to radial instability rate. At large scaled impact, the limiting constitutive deflection follows an inverse fourth power law. A nonzero fixed mass introduces a slower inverse-impact contribution, producing a second nonuniform limit. Equality of these two terms becomes a physical returning-ray scale only above the critical impact, and a useful weak-bending approximation additionally requires sufficient separation from the barrier.

Independent representations of the energy integral, stationary orbit and scattering angle support these distinctions. The numerical comparisons quantify both the gain from complete second order and the parameter dependence of its useful range. They also separate accurate evaluation of an exact ray from the error of an asymptotic description. The resulting construction links admissibility, mass normalization and exterior scattering within one specified limiting problem.

A natural extension is to apply the same separation of scales to other constitutive responses and coupled parameter paths. Comparing their limiting barriers and correction rates would identify which features depend on the fractional model and which persist across a broader class of admissible electromagnetic exteriors.

\section*{Acknowledgments}
Edson Otoniel acknowledges support from Funda\c{c}\~ao Cearense de Apoio ao Desenvolvimento Cient\'ifico e Tecnol\'ogico through grant BP6-0241-00335.01.00/25. R. Martínez von Dossow acknowledges financial support from the Agencia Nacional de Investigación y Desarrollo (ANID), Chile, through FONDECYT Postdoctorado, Grant No. 3260624.

\afterpage{\balance}
\bibliographystyle{elsarticle-num}
\bibliography{ref}

@article{Kruglov,
  author = {S.I. Kruglov},
  title = {{Born--Infeld-type electrodynamics and magnetic black holes}},
  journal = {Annals of Physics},
  year = {2017},
  doi = {10.1016/j.aop.2017.06.008},
  volume = {383},
  pages = {550-559},
  eprint = {1707.04495},
  archivePrefix = {arXiv},
  primaryClass = {gr-qc},
  note = {Corrigendum: Annals Phys. 434 (2021) 168625, doi:10.1016/j.aop.2021.168625},
}

@article{Novello,
  author = {M. Novello and V. A. De Lorenci and J. M. Salim and R. Klippert},
  title = {{Geometrical aspects of light propagation in nonlinear electrodynamics}},
  journal = {Physical Review D},
  year = {2000},
  doi = {10.1103/physrevd.61.045001},
  volume = {61},
  number = {4},
  pages = {045001},
  eprint = {gr-qc/9911085},
  archivePrefix = {arXiv},
  primaryClass = {gr-qc},
}

@article{He2022,
  author = {Aoyun He and Jun Tao and Peng Wang and Yadong Xue and Lingkai Zhang},
  title = {{Effects of Born--Infeld electrodynamics on black hole shadows}},
  journal = {The European Physical Journal C},
  year = {2022},
  doi = {10.1140/epjc/s10052-022-10637-x},
  volume = {82},
  number = {8},
  pages = {683},
  eprint = {2205.12779},
  archivePrefix = {arXiv},
  primaryClass = {gr-qc},
}

@article{Habibina,
  author = {A. S. Habibina and H. S. Ramadhan},
  title = {{Geodesic of nonlinear electrodynamics and stable photon orbits}},
  journal = {Physical Review D},
  year = {2020},
  doi = {10.1103/physrevd.101.124036},
  volume = {101},
  number = {12},
  pages = {124036},
  eprint = {2007.03211},
  archivePrefix = {arXiv},
  primaryClass = {gr-qc},
}

@misc{Ramadhan,
  author = {Ramadhan, H. S. and Fauzi, M. F. and Witjaksana, D. A. and Sulaksono, A.},
  title = {{Photon Propagation and Black Hole Imaging in Kruglov Nonlinear Electrodynamics}},
  year = {2026},
  url = {https://arxiv.org/abs/2604.22309v3},
  eprint = {2604.22309},
  archivePrefix = {arXiv},
  primaryClass = {gr-qc},
}

@article{Gralla2019,
  author = {Samuel E. Gralla and Daniel E. Holz and Robert M. Wald},
  title = {{Black hole shadows, photon rings, and lensing rings}},
  journal = {Physical Review D},
  year = {2019},
  doi = {10.1103/physrevd.100.024018},
  volume = {100},
  number = {2},
  pages = {024018},
  eprint = {1906.00873},
  archivePrefix = {arXiv},
  primaryClass = {astro-ph.HE},
}

@article{Bozza2002,
  author = {V. Bozza},
  title = {{Gravitational lensing in the strong field limit}},
  journal = {Physical Review D},
  year = {2002},
  doi = {10.1103/physrevd.66.103001},
  volume = {66},
  number = {10},
  pages = {103001},
  eprint = {gr-qc/0208075},
  archivePrefix = {arXiv},
  primaryClass = {gr-qc},
}

@article{Tsukamoto2017,
  author = {Naoki Tsukamoto},
  title = {{Deflection angle in the strong deflection limit in a general asymptotically flat, static, spherically symmetric spacetime}},
  journal = {Physical Review D},
  year = {2017},
  doi = {10.1103/physrevd.95.064035},
  volume = {95},
  number = {6},
  pages = {064035},
  eprint = {1612.08251},
  archivePrefix = {arXiv},
  primaryClass = {gr-qc},
}

@article{BornInfeld1934,
  author = {Max Born and L. Infeld},
  title = {{Foundations of the new field theory}},
  journal = {Proc. R. Soc. Lond. A},
  year = {1934},
  doi = {10.1098/rspa.1934.0059},
  volume = {144},
  number = {852},
  pages = {425-451},
}

@article{Boillat1970,
  author = {Guy Boillat},
  title = {{Nonlinear Electrodynamics: Lagrangians and Equations of Motion}},
  journal = {Journal of Mathematical Physics},
  year = {1970},
  doi = {10.1063/1.1665231},
  volume = {11},
  number = {3},
  pages = {941-951},
}

@article{KruglovNotes,
  author = {S. I. Kruglov},
  title = {{Notes on Born--Infeld-type electrodynamics}},
  journal = {Modern Physics Letters A},
  year = {2017},
  doi = {10.1142/s0217732317502017},
  volume = {32},
  number = {36},
  pages = {1750201},
  eprint = {1612.04195},
  archivePrefix = {arXiv},
  primaryClass = {physics.gen-ph},
}

@article{Abalos2015,
  author = {Fernando Abalos and Federico Carrasco and {\'E}rico Goulart and Oscar Reula},
  title = {{Nonlinear electrodynamics as a symmetric hyperbolic system}},
  journal = {Physical Review D},
  year = {2015},
  doi = {10.1103/physrevd.92.084024},
  volume = {92},
  number = {8},
  pages = {084024},
  eprint = {1507.02262},
  archivePrefix = {arXiv},
  primaryClass = {gr-qc},
}

@article{ObukhovRubilar2002,
  author = {Yuri N. Obukhov and Guillermo F. Rubilar},
  title = {{Fresnel analysis of wave propagation in nonlinear electrodynamics}},
  journal = {Physical Review D},
  year = {2002},
  doi = {10.1103/physrevd.66.024042},
  volume = {66},
  number = {2},
  pages = {024042},
  eprint = {gr-qc/0204028},
  archivePrefix = {arXiv},
  primaryClass = {gr-qc},
}

@article{Bronnikov2001,
  author = {K. A. Bronnikov},
  title = {{Regular magnetic black holes and monopoles from nonlinear electrodynamics}},
  journal = {Physical Review D},
  year = {2001},
  doi = {10.1103/physrevd.63.044005},
  volume = {63},
  number = {4},
  pages = {044005},
  eprint = {gr-qc/0006014},
  archivePrefix = {arXiv},
  primaryClass = {gr-qc},
}

@article{ShabadUsov2011,
  author = {Anatoly E. Shabad and Vladimir V. Usov},
  title = {{Effective Lagrangian in nonlinear electrodynamics and its properties of causality and unitarity}},
  journal = {Physical Review D},
  year = {2011},
  doi = {10.1103/physrevd.83.105006},
  volume = {83},
  number = {10},
  pages = {105006},
  eprint = {1101.2343},
  archivePrefix = {arXiv},
  primaryClass = {hep-th},
}

@article{GibbonsHerdeiro2001,
  author = {G. W. Gibbons and C. A. R. Herdeiro},
  title = {{Born-Infeld theory and stringy causality}},
  journal = {Physical Review D},
  year = {2001},
  doi = {10.1103/physrevd.63.064006},
  volume = {63},
  number = {6},
  pages = {064006},
  eprint = {hep-th/0008052},
  archivePrefix = {arXiv},
  primaryClass = {hep-th},
}

@article{Cardoso2009,
  author = {Vitor Cardoso and Alex S. Miranda and Emanuele Berti and Helvi Witek and Vilson T. Zanchin},
  title = {{Geodesic stability, Lyapunov exponents, and quasinormal modes}},
  journal = {Physical Review D},
  year = {2009},
  doi = {10.1103/physrevd.79.064016},
  volume = {79},
  number = {6},
  pages = {064016},
  eprint = {0812.1806},
  archivePrefix = {arXiv},
  primaryClass = {hep-th},
}

@article{Stefanov2010,
  author = {Ivan Zh. Stefanov and Stoytcho S. Yazadjiev and Galin G. Gyulchev},
  title = {{Connection between Black-Hole Quasinormal Modes and Lensing in the Strong Deflection Limit}},
  journal = {Physical Review Letters},
  year = {2010},
  doi = {10.1103/physrevlett.104.251103},
  volume = {104},
  number = {25},
  pages = {251103},
  eprint = {1003.1609},
  archivePrefix = {arXiv},
  primaryClass = {gr-qc},
}

@article{BornAgain2024,
  author = {Jorge G. Russo and Paul K. Townsend},
  title = {{Born again}},
  journal = {SciPost Physics},
  year = {2024},
  doi = {10.21468/scipostphys.16.5.124},
  volume = {16},
  number = {5},
  pages = {124},
  eprint = {2401.04167},
  archivePrefix = {arXiv},
  primaryClass = {hep-th},
}

@article{AyonBeatoGarcia1998,
  author = {Eloy Ay{\'o}n-Beato and Alberto Garc{\'i}a},
  title = {{Regular Black Hole in General Relativity Coupled to Nonlinear Electrodynamics}},
  journal = {Physical Review Letters},
  year = {1998},
  volume = {80},
  number = {23},
  pages = {5056-5059},
  doi = {10.1103/physrevlett.80.5056},
  eprint = {gr-qc/9911046},
  archivePrefix = {arXiv},
}

@article{AyonBeatoGarcia2000,
  author = {Eloy Ay{\'o}n-Beato and Alberto Garc{\'i}a},
  title = {{The Bardeen model as a nonlinear magnetic monopole}},
  journal = {Physics Letters B},
  year = {2000},
  volume = {493},
  number = {1-2},
  pages = {149-152},
  doi = {10.1016/s0370-2693(00)01125-4},
  eprint = {gr-qc/0009077},
  archivePrefix = {arXiv},
}

@article{Dymnikova2004,
  author = {Irina Dymnikova},
  title = {{Regular electrically charged vacuum structures with de Sitter centre in nonlinear electrodynamics coupled to general relativity}},
  journal = {Classical and Quantum Gravity},
  year = {2004},
  volume = {21},
  number = {18},
  pages = {4417-4428},
  doi = {10.1088/0264-9381/21/18/009},
  eprint = {gr-qc/0407072},
  archivePrefix = {arXiv},
}

@article{MorenoSarbach2003,
  author = {Claudia Moreno and Olivier Sarbach},
  title = {{Stability properties of black holes in self-gravitating nonlinear electrodynamics}},
  journal = {Physical Review D},
  year = {2003},
  volume = {67},
  number = {2},
  pages = {024028},
  doi = {10.1103/physrevd.67.024028},
  eprint = {gr-qc/0208090},
  archivePrefix = {arXiv},
}

@article{CunhaHerdeiro2018,
  author = {Pedro V. P. Cunha and Carlos A. R. Herdeiro},
  title = {{Shadows and strong gravitational lensing: a brief review}},
  journal = {General Relativity and Gravitation},
  year = {2018},
  volume = {50},
  number = {4},
  pages = {42},
  doi = {10.1007/s10714-018-2361-9},
  eprint = {1801.00860},
  archivePrefix = {arXiv},
}

@misc{Hui2025,
  author = {Hui, Siyuan and Mu, Benrong and Wang, Peng},
  title = {{Gravitational Lensing by Black Holes in Einstein-nonlinear Electrodynamic Theories with Multiple Photon Spheres}},
  year = {2025},
  url = {https://arxiv.org/abs/2507.13048},
  eprint = {2507.13048},
  archivePrefix = {arXiv},
}

@article{PerlickTsupko2022,
  author = {Volker Perlick and Oleg Yu. Tsupko},
  title = {{Calculating black hole shadows: Review of analytical studies}},
  journal = {Physics Reports},
  year = {2022},
  volume = {947},
  pages = {1-39},
  doi = {10.1016/j.physrep.2021.10.004},
  eprint = {2105.07101},
  archivePrefix = {arXiv},
}

@article{GibbonsWerner2008,
  author = {G W Gibbons and M C Werner},
  title = {{Applications of the Gauss--Bonnet theorem to gravitational lensing}},
  journal = {Classical and Quantum Gravity},
  year = {2008},
  volume = {25},
  number = {23},
  pages = {235009},
  doi = {10.1088/0264-9381/25/23/235009},
  eprint = {0807.0854},
  archivePrefix = {arXiv},
}

@article{Claudel2001,
  author = {Clarissa-Marie Claudel and K. S. Virbhadra and G. F. R. Ellis},
  title = {{The geometry of photon surfaces}},
  journal = {Journal of Mathematical Physics},
  year = {2001},
  volume = {42},
  number = {2},
  pages = {818-838},
  doi = {10.1063/1.1308507},
  eprint = {gr-qc/0005050},
  archivePrefix = {arXiv},
}

@article{VirbhadraEllis2000,
  author = {K. S. Virbhadra and George F. R. Ellis},
  title = {{Schwarzschild black hole lensing}},
  journal = {Physical Review D},
  year = {2000},
  volume = {62},
  number = {8},
  pages = {084003},
  doi = {10.1103/physrevd.62.084003},
  eprint = {astro-ph/9904193},
  archivePrefix = {arXiv},
}

@article{Perlick2004,
  author = {Volker Perlick},
  title = {{Exact gravitational lens equation in spherically symmetric and static spacetimes}},
  journal = {Physical Review D},
  year = {2004},
  volume = {69},
  number = {6},
  pages = {064017},
  doi = {10.1103/physrevd.69.064017},
  eprint = {gr-qc/0307072},
  archivePrefix = {arXiv},
}
\end{document}